%% file: main.tex
\documentclass[aps,physrev,reprint,noeprint,superscriptaddress]{revtex4-2} 
\usepackage{bm, braket,amsmath,mathtools,comment}
\usepackage{graphicx,color}
\usepackage[unicode]{hyperref}
\usepackage[normalem]{ulem}

\begin{document}
\title{Evaluating thermal expectation values by almost ideal sampling with Trotter gates}
\author{Shimpei Goto}
\email[]{goto.las@tmd.ac.jp}
\affiliation{College of Liberal Arts and Sciences, Tokyo Medical and Dental University, Ichikawa, Chiba 272-0827, Japan}
\author{Ryui Kaneko}
\email[]{rkaneko@phys.kindai.ac.jp}
\author{Ippei Danshita}
\email[]{danshita@phys.kindai.ac.jp}
\affiliation{Department of Physics, Kindai University, Higashi-Osaka city, Osaka 577-8502, Japan}
\date{\today}
\begin{abstract}
    We investigate the sampling efficiency for the simulations of quantum many-body systems at finite temperatures when initial sampling states are generated by applying Trotter gates to random phase product states (RPPSs).
    We restrict the number of applications of Trotter gates to be proportional to the system size, and thus the preparation would be easily accomplished in fault-tolerant quantum computers.
    When the Trotter gates are made from a nonintegrable Hamiltonian, we observe that the sampling efficiency increases with system size.
    This trend means that almost ideal sampling of initial states can be achieved in sufficiently large systems.
    We also find that the sampling efficiency is almost equal to that obtained by a typical pure quantum (TPQ) state method utilizing Haar random sampling in some cases.
    These findings suggest that chaotic Hamiltonian dynamics can transform RPPSs into an alternative to TPQ states for evaluating thermal expectation values.
\end{abstract}
\maketitle
\section{Introduction\label{sec:introduction}}

Stimulated by the rapid development of quantum information technology, recent studies have proposed many new applications of quantum computers.
Techniques for simulating nonunitary imaginary-time evolution on quantum circuits, which are based on variational approaches or nonunitary measurements~\cite{mcardle_variational_2019,motta_determining_2020,lin_real-_2021,sun_quantum_2021,liu_probabilistic_2021,nishi_implementation_2021,kosugi_imaginary-time_2022}, are one of such applications.
An important outcome of imaginary-time evolution is the simulation of quantum many-body systems at finite temperatures, whose large-scale classical simulations are available only in some limited classes such as low-dimensional systems~\cite{feiguin_finite-temperature_2005,white_minimally_2009,stoudenmire_minimally_2010,iitaka_random_2020,goto_matrix_2021-1,iwaki_thermal_2021} and those that are free from the negative sign problem~\cite{gubernatis_quantum_2016}. 

One of the methods for simulating quantum many-body systems at finite temperatures has been indeed tested on quantum devices for systems with four spins in Ref.~\cite{sun_quantum_2021}.
There the thermal average of an observable has been evaluated by summing up the full trace over the entire Hilbert space of the system or randomly sampling certain types of product states as initial states for the imaginary-time evolution.
The former approach cannot be applied to systems with hundreds of spins because of the exponential growth of the Hilbert space size.
Moreover, it is known in the context of classical simulations of quantum many-body systems at finite temperatures that the latter approach cannot be applied to those in a practical sense, either, because of poor efficiency of the sampling~\cite{goto_matrix_2021-1}. 
In terms of the sampling efficiency, Haar random states are ideal in the sense that sampling dependence decreases exponentially with the system size~\cite{sugiura_thermal_2012,sugiura_canonical_2013}.
The classical algorithm based on this fact is often referred to as a typical pure quantum (TPQ) state method.
The execution of the large-scale TPQ state method is, however, difficult even on quantum computers because exponentially many random numbers are required for the preparation of initial Haar random states.

This situation motivates us to introduce a more efficient and applicable sampling method for evaluating the thermal average of an observable.
For this purpose, we propose a sampling method for evaluating the thermal average, in which we use highly entangled states created by applying Trotter gates to random-phase product states (RPPSs) as initial states for the imaginary-time evolution.
Performing numerical simulations on classical computers with available system sizes, we investigate the sampling efficiency in various spin models at one spatial dimension.
We find that the choice of the Trotter gates significantly affects the sampling efficiency. When Trotter gates are made from nonintegrable Hamiltonians, we observe that the sampling efficiency increases with the system size, just as in the TPQ state method.
By contrast, we observe that this increase is absent when Trotter gates are made from an integrable Hamiltonian.
In particular, when Trotter gates are prepared from the mixed-field Ising model, the efficiency of the proposed sampling scheme is almost equal to that of the TPQ state method as for XXZ and mixed-field Ising models, suggesting that the ideal sampling is achieved.
In our proposed method, the depth of the operations of the Trotter gates scales linearly with the system size \(L\) so that the proposed initial states are available on quantum computers.

These results show that our proposed method provides a way to construct an alternative to Haar random states for evaluating thermal expectation values on quantum computers.
Similar attempts based on unitary--\(t\) designs have been proposed recently~\cite{seki_energy-filtered_2022-1, coopmans_predicting_2022}.
In this paper, we find another possibility to prepare the alternative: Increasing the sampling efficiency of the RPPS-based approach equivalent to that of the TPQ method with the help of chaotic Hamiltonian dynamics.

The rest of the paper is organized as follows: In Sec.~\ref{sec:methods}, we explain the details of how to evaluate thermal expectation values in the simulations.
In Sec.~\ref{sec:efficiency}, the efficiency of sampling is introduced.
The results of numerical simulations on spin-1/2 chains are given in Sec.~\ref{sec:results}.
In Sec.~\ref{sec:summaries}, we discuss how the choice of the Trotter gates improves sampling efficiency and how to evaluate the performance of Trotter gates without accessing the norms of wavefunctions in quantum simulations. Summary is also given in this section.

\section{How to evaluate thermal expectation values\label{sec:methods}}
The objective of the simulation is to evaluate thermal expectation values
\begin{align}
    \label{eq:th_exp}
    \braket{\hat{O}}_\beta = \frac{\mathrm{Tr} \left [\hat{O} e^{-\beta \hat{H}}\right ]}{\mathrm{Tr}e^{-\beta \hat{H}}}.
\end{align}
Here, \(\hat{O}\) is an observable operator that one would like to evaluate, \(\beta \) is the inverse temperature, and \(\hat{H}\) denotes the Hamiltonian of the system.
To evaluate the quantity in Eq.~\eqref{eq:th_exp}, the evaluation of the trace of operators are required.
We evaluate its value from the sampling of pure states. 

The trace of an operator can be obtained from the sampling average of the expectation value from some class of random pure states, i.e., 
\begin{align}
    \mathrm{Tr} \hat{O} = C \overline{\braket{\psi|\hat{O}|\psi}},
\end{align}
where the overline means the expectation value over random states \(\ket{\psi}\).
The prefactor \(C\) only depends on the choice of the class and the size of the Hilbert space.
With using imaginary-time evolution techniques, one can estimate \(\mathrm{Tr}\left [ \hat{O} e^{-\beta \hat{H}}\right ] = C \overline{\braket{\psi|e^{-\beta \hat{H}/2}\hat{O}e^{-\beta \hat{H}/2}|\psi}}\) and \(\mathrm{Tr} e^{-\beta \hat{H}} = C \overline{\braket{\psi|e^{-\beta \hat{H}/2}e^{-\beta \hat{H}/2}|\psi}}\)~\cite{sun_quantum_2021}.
The ratio of the two estimations gives the thermal expectation value.

The Haar random states, \(\ket{\psi_\mathrm{Haar}} = \sum_i c_i \ket{i}\), are frequently used as such random pure states~\cite{jin_random_2020}.
Here, \(\ket{i}\) is an orthonormal basis of the Hilbert space and coefficients \(c_i\) are complex numbers drawn uniformly from the unit sphere \(\sum_i|c_i|^2 = 1\).
The states possess entanglement entropy that follows a volume-law scaling.
The nice feature of the Haar random states is that expectation values, which are evaluated with the states prepared with the imaginary-time evolution of the initial Haar random states, are independent of the samples in the thermodynamic limit.
\textcite{sugiura_thermal_2012,sugiura_canonical_2013} have shown that the fluctuations of expectation values decrease exponentially with system size.
The drawback is the difficulty in its preparation.
Exponentially many random numbers are required so that the preparation would be difficult even on quantum computers.
To avoid this difficulty, an approach using the unitary--2~\cite{seki_energy-filtered_2022-1} or 3~\cite{coopmans_predicting_2022} design instead of the Haar random states has been proposed recently.

\textcite{iitaka_random_2004} have proposed other class of (unnormalized) random states, random phase states \(\ket{\psi_\mathrm{RP}} = \sum_i \mathrm{e}^{i \theta_i} \ket{i}\).
Here, \(\theta_i\) is a random real number drawn uniformly from \([0, 2\pi )\).
These states also require exponentially many random numbers and thus possess volume-law entanglement.
Recently, \textcite{iitaka_random_2020} proposes a variant class of random phase states, i.e., RPPSs
\begin{align}
    \ket{\psi_\mathrm{RPP}} = \bigotimes_i \sum_\sigma \left (e^{i \theta_{\sigma_i}}\ket{\sigma_i} \right )
\end{align}
to bring the idea of random phase states to matrix-product-state simulations.
Here, \(\ket{\sigma_i}\) is an orthonormal basis of the local Hilbert at site \(i\) and a phase \(\theta_{\sigma_i}\) is also drawn uniformly from \([0, 2\pi )\).
The RPPSs are characterized by random numbers whose number grows only linearly with system size.
Since RPPSs are product states, the preparation of each RPPS is accomplished by local operations.
Therefore, duplication of RPPSs on separate quantum computers is an easy task and it is beneficial for estimating standard expectation values which also requires sampling on quantum computers.

However, the authors of the current paper have found in their previous work~\cite{goto_matrix_2021-1} that the sampling based on RPPSs is quite inefficient even on the simulations of the simple Heisenberg chain.
To get rid of this inefficiency, the authors have proposed to make initial RPPSs slightly entangled by applying Trotter gates,
\begin{align}
    \ket{\psi_\mathrm{ini}} = \hat{U}^n_\mathrm{T}(\tau)\bigotimes_i \sum_\sigma \left (e^{i \theta_{\sigma_i}}\ket{\sigma_i} \right ).
\end{align}
Here, \(n\) is the number of operations which is restricted to be proportional to system size in this study, \(\tau \) is the time step for the Trotter gates, \(\hat{U}_\mathrm{T}(\tau) = e^{-i\tau \hat{H}_\mathrm{odd}}e^{-i\tau \hat{H}_\mathrm{even}}\) is the first-order Trotter gate obtained from the Hamiltonian \(\hat{H}_\mathrm{T} = \hat{H}_\mathrm{odd} + \hat{H}_\mathrm{even}\), where \(\hat{H}_\mathrm{odd}\) (\(\hat{H}_\mathrm{even}\)) acts only on sites \(2i-1\) and \(2i\) (\(2i\) and \(2i+1\)).
This approach succeeds in improving the efficiency of the sampling approach.
The Trotter Hamiltonian \(\hat{H}_\mathrm{T}\) should be different from the system Hamiltonian \(\hat{H}\) since the effects of the Trotter gates are absent if \(\hat{U}_\mathrm{T}(\tau)\) and \(\hat{H}\) commute.
In Ref.~\cite{goto_matrix_2021-1}, the number of operations \(n\) is restricted up to ten in order not to increase excessively the entanglement entropy of wavefunctions because this approach is designed for classical computers.
If one can perform this approach on quantum computers, such restriction is unnecessary.
Hereafter we investigate how efficient this approach can become without the restriction.

\section{Sampling efficiency\label{sec:efficiency}}
The thermal expectation value estimated from \(M\) samples can be expresses as
\begin{align}
    \label{eq:th_exp_sample}
    \braket{\hat{O}}_\beta = \sum^M_{m=1} \frac{\braket{\psi_m|e^{-\frac{\beta}{2} \hat{H}}\hat{O}e^{-\frac{\beta}{2} \hat{H}}|\psi_m}}{\braket{\psi_m|e^{-\beta \hat{H}}|\psi_m}}\frac{\braket{\psi_m|e^{-\beta \hat{H}}|\psi_m}}{\sum^M_{l=1}\braket{\psi_l|e^{-\beta \hat{H}}|\psi_l}}.
\end{align}
Here, \(\ket{\psi_m}\) is the \(m\)th random initial state.
From this expression, the weight of \(m\)th sample can be defined as
\begin{align}
    w_m = \frac{\braket{\psi_m|e^{-\beta \hat{H}}|\psi_m}}{\sum^M_{l=1}\braket{\psi_l|e^{-\beta \hat{H}}|\psi_l}}
\end{align}
since \(w_m > 0\) and \(\sum_{m} w_m = 1\) hold.
A thermal expectation value estimated from sampling~\eqref{eq:th_exp_sample} can be regarded as a weighted sum, and the distribution of weights should be uniform for efficient sampling.
With evaluating the fluctuation of weights by the information entropy \(I = -\sum^M_{m=1}w_m \ln w_m\), the uniform weights give \(I = \ln M\).
Consequently, the efficiency of sampling can be introduced as~\cite{goto_matrix_2021-1}
\begin{align}
    \label{eq:efficiency}
    \eta = \frac{e^I}{M}.
\end{align}
This quantity relates to the fluctuation of the partition function \(\mathrm{Tr}e^{-\beta \hat{H}}\)~\cite{iwaki_purity_2022-1}.
If \(\eta \) is sufficiently close to unity, the norm of a single sampled state after the imaginary-time evolution gives a good estimation of the partition function, i.e., a single pure state possesses the information of finite-temperature quantities given by the Gibbs state.
Moreover, in such a situation, \(w_i\) is almost given by \(1/M\)  and a thermal expectation value can be obtained approximately by
\begin{align}
    \label{eq:th_exp_simple}
    \braket{\hat{O}}^\prime_\beta = \frac{1}{M}\sum^{M}_{m=1}\frac{\braket{\psi_m|e^{-\frac{\beta}{2} \hat{H}}\hat{O}e^{-\frac{\beta}{2} \hat{H}}|\psi_m}}{\braket{\psi_m|e^{-\beta \hat{H}}|\psi_m}},
\end{align}
i.e., the sample average of standard expectation values given by states normalized after the imaginary-time evolution.
This fact is useful for simulations on quantum computers since the evaluation of the norms of imaginary-time evolved states, which would be troublesome because of its exponential dependence on the inverse temperature, can be skipped.

When initial states are Haar random states, i.e., one employs the TPQ state method, the fluctuation of the norms of imaginary-time evolved states decreases exponentially with increasing system size~\cite{sugiura_thermal_2012,sugiura_canonical_2013}.
Consequently, the sampling efficiency \(\eta \) increases with system size and the ideal sampling can be achieved in sufficiently large systems.
The system-size dependence of the sampling efficiency is an important character of sampling scheme.

\section{System-size dependencies of sampling efficiency\label{sec:results}}
First, we consider a case where the system Hamiltonian is given by the Heisenberg chain,
\begin{align}
    \hat{H} = J \sum^{L-1}_{i=1} (\hat{S}^x_i \hat{S}^x_{i+1} + \hat{S}^y_i \hat{S}^y_{i+1} + \hat{S}^z_i \hat{S}^z_{i+1}),
\end{align}
and the Trotter Hamiltonian is given by the XXZ chain with staggered magnetic field,
\begin{eqnarray}
    \hat{H}_\mathrm{T} &=& J \sum^{L-1}_{i=1} (\hat{S}^x_i \hat{S}^x_{i+1} + \hat{S}^y_i \hat{S}^y_{i+1} 
    + \Delta \hat{S}^z_i \hat{S}^z_{i+1})
    \nonumber \\
    && + h\sum^L_{i=1} {(-1)}^i \hat{S}^z_i.
\end{eqnarray}
Here, \(J\) is the strength of the spin-spin exchange interaction, \(\hat{S}^x_i, \hat{S}^y_i\), and \(\hat{S}^z_i\) are spin-1/2 operators at site \(i\), \(L\) is the number of lattice sites, \(\Delta \) denotes the anisotropy parameter, which we set to \(5.0\) in this study, and \(h\) is the strength of the staggered magnetic field.
It should be noted that the Trotter Hamiltonian \(\hat{H}_\mathrm{T}\) is nonintegrable (integrable) when \(h\) is finite (zero)~\cite{brenes_high-temperature_2018}.
We randomly choose RPPSs and operate Trotter gates made from \(\hat{H}_\mathrm{T}\) \(2L\) times to generate volume-law entangled states.
The time step of the Trotter gates \(\tau \) is set to \(10.0 J^{-1}\).
Since the volume-law entangled states emerge in simulations, we prepare the total Hilbert space and restrict \(L\) up to 20.
Imaginary-time evolution is performed with the Al-Mohy-Higham method, which is the Taylor expansion based method with error analysis~\cite{al-mohy_computing_2011}. 

\begin{figure}
    \includegraphics[width=0.95\linewidth]{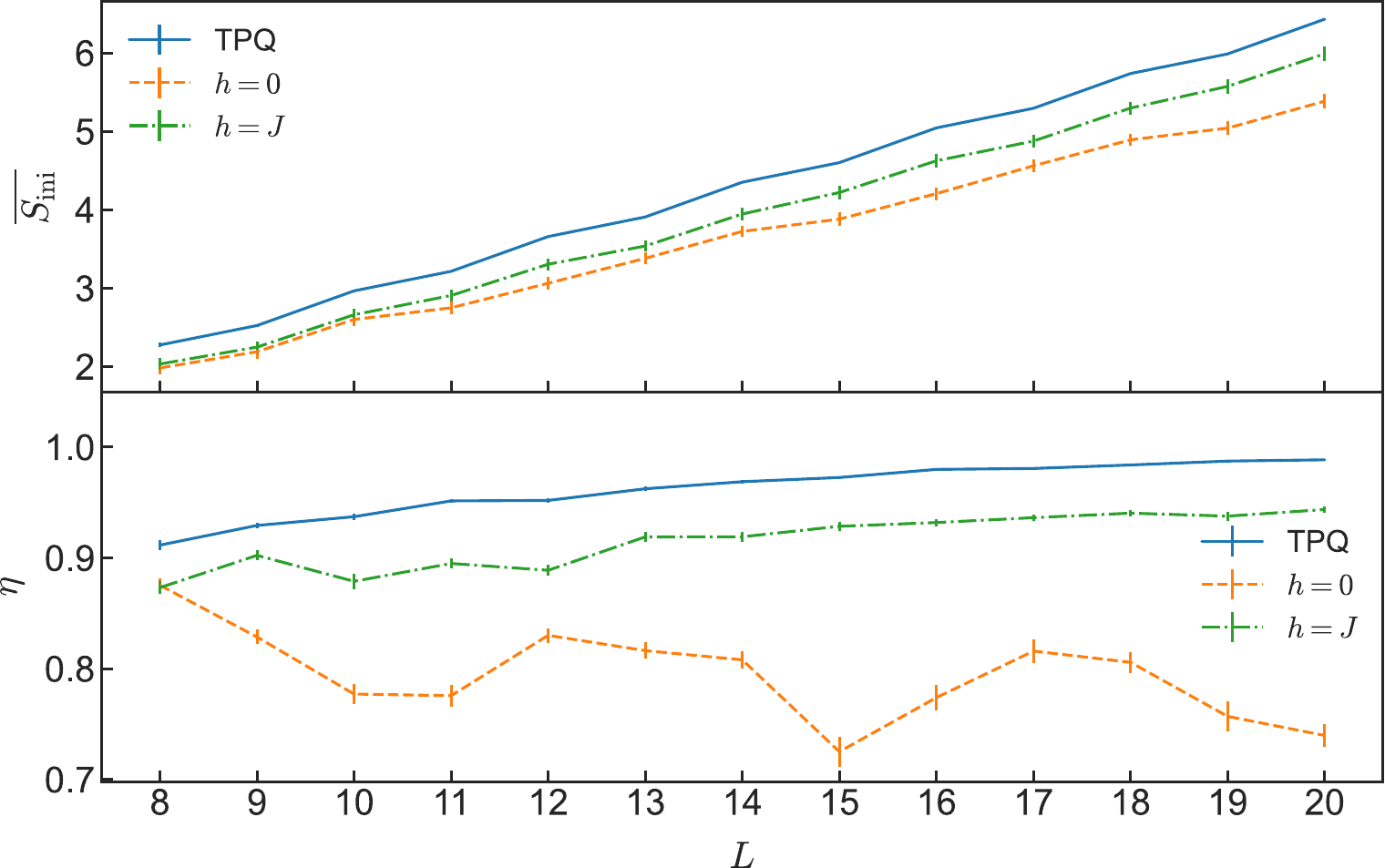}
    \caption{System-size dependence of the average entanglement entropy of sampled initial states (top panel) and the efficiency (bottom panel) obtained from 1024 samples for the TPQ states (blue solid line), the RPPSs evolved by the Trotter gates obtained from the XXZ chain without the staggered magnetic field (orange dashed line), and those evolved by the Trotter gates obtained from the XXZ chain with the staggered magnetic field (green dashed-dotted line). The inverse temperature \(\beta \) is set to \(3.0 J^{-1}\). The error bars indicate 1\(\sigma \) uncertainty. The 1\(\sigma \) uncertainty of the efficiency is estimated from the bootstrap analysis with 4000 resampled data.\label{fig:HeisXXZ}}
\end{figure}

\begin{figure}
    \includegraphics[width=0.95\linewidth]{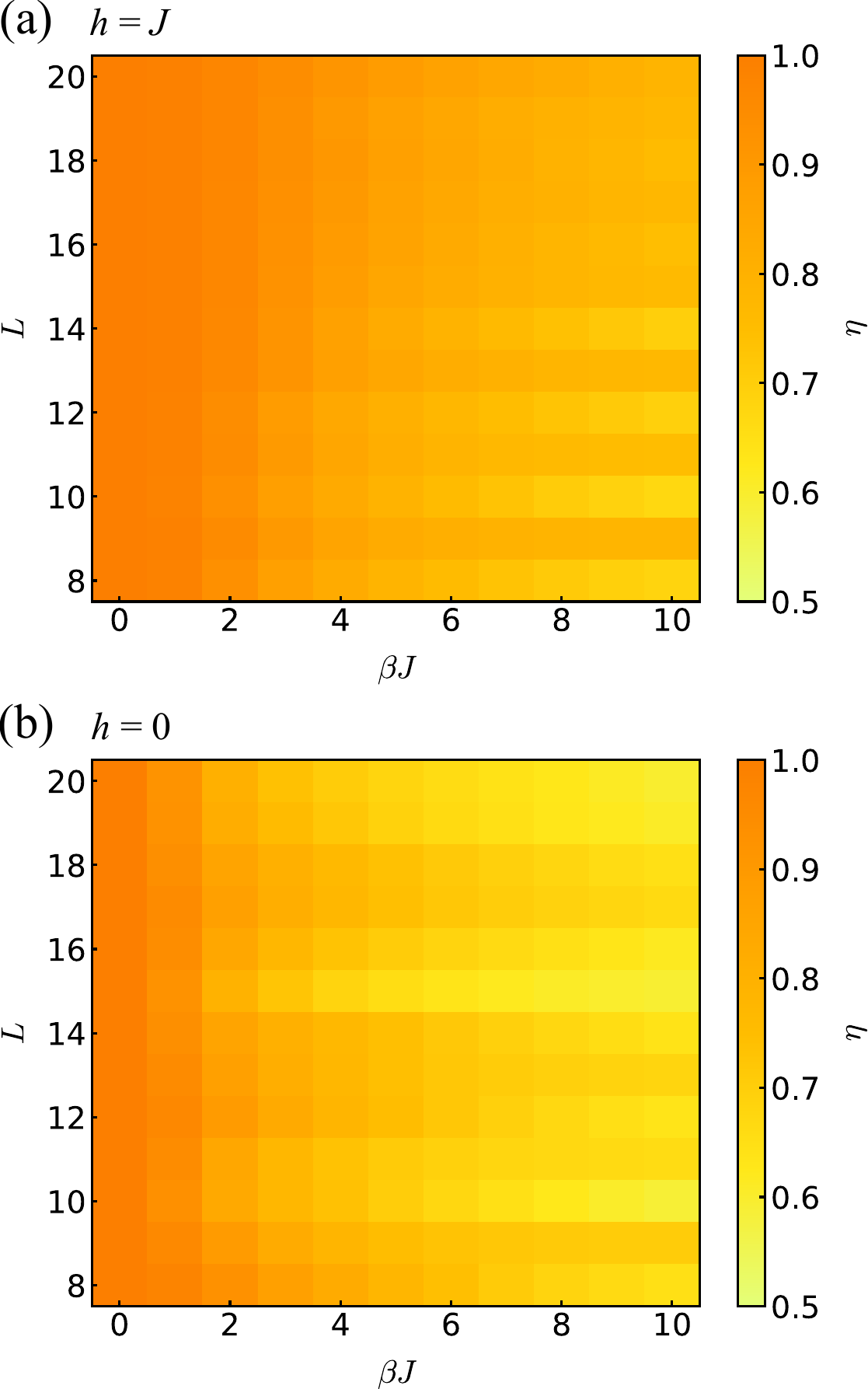}
    \caption{System-size and inverse temperature dependence of the efficiency obtained from 1024 samples for the RPPSs evolved by the Trotter gates obtained from the XXZ chain with (a) and without (b) the staggered magnetic field.\label{fig:eff_map}}
\end{figure}

Figure~\ref{fig:HeisXXZ} represents the system-size dependencies of the averaged entanglement entropy of initial states and the efficiency that is estimated from 1024 samples of the TPQ states and the RPPSs that are evolved by the Trotter gates obtained from the XXZ chain with and without the staggered magnetic field.
The entanglement entropy is defined as \(S_\mathrm{ini} = -\mathrm{Tr}\hat{\rho}_A \ln \hat{\rho}_A\), where \(\hat{\rho}_A\) is the reduced density matrix \(\hat{\rho}_A = \mathrm{Tr}_B \ket{\psi_\mathrm{ini}}\bra{\psi_\mathrm{ini}}\) and \(\ket{\psi_\mathrm{ini}}\) is an initial state for the imaginary-time evolution prepared by the applications of the Trotter gates.
Here, the subsystem \(B\) is the right half (\(i > \lfloor L/2 \rfloor \)) of the system.
In every approach, the initial entanglement entropy increases linearly with the system sizes: The initial states possess volume-law entanglement.
On the contrary, the efficiencies of the RPPSs with and without the staggered magnetic field show different system-size dependencies.
With the staggered magnetic field, the efficiency gradually increases with the system size like the TPQ state method.
Such an increase is, however, absent without the staggered field.
These trends are observed in the whole temperature region investigated in this study, as demonstrated in Fig.~\ref{fig:eff_map}.
As stated in the above, the presence of the staggered field affects the integrability of the Trotter Hamiltonian.
From the above results, we conjecture that the efficiency \(\eta \) asymptotically approaches unity for the Trotter gates made from a nonintegrable Hamiltonian while it does not for those made from an integrable Hamiltonian.

\begin{figure}
    \includegraphics[width=0.95\linewidth]{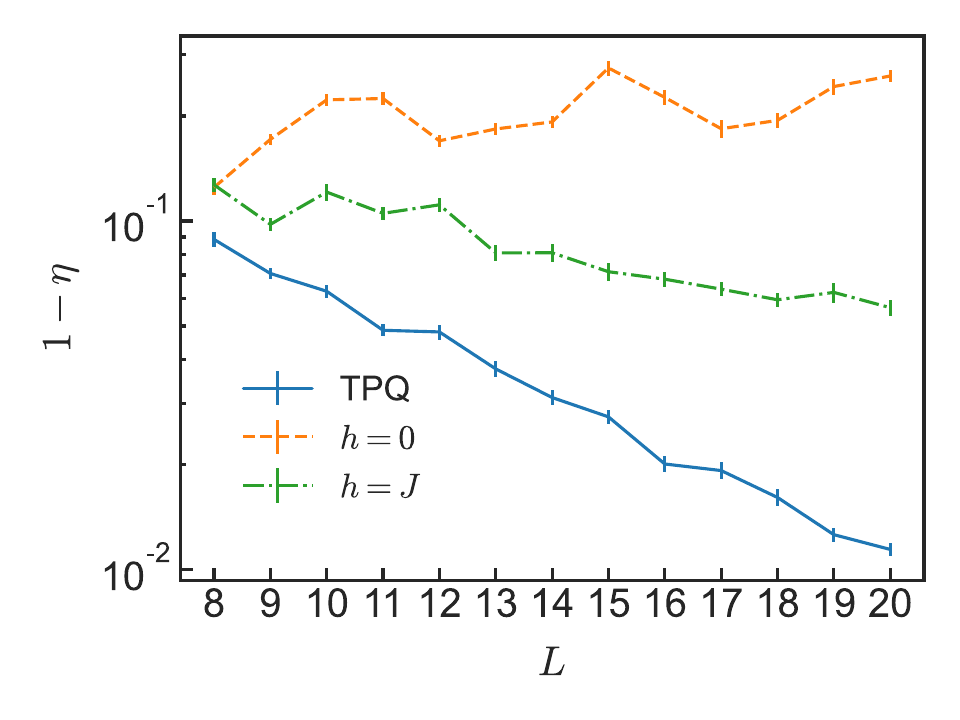}
    \caption{System-size dependence of the difference of the efficiency from unity. The efficiencies are the same with those shown in the bottom panel of Fig.~\ref{fig:HeisXXZ}.\label{fig:eff_log}}
\end{figure}

\begin{figure}
    \includegraphics[width=0.95\linewidth]{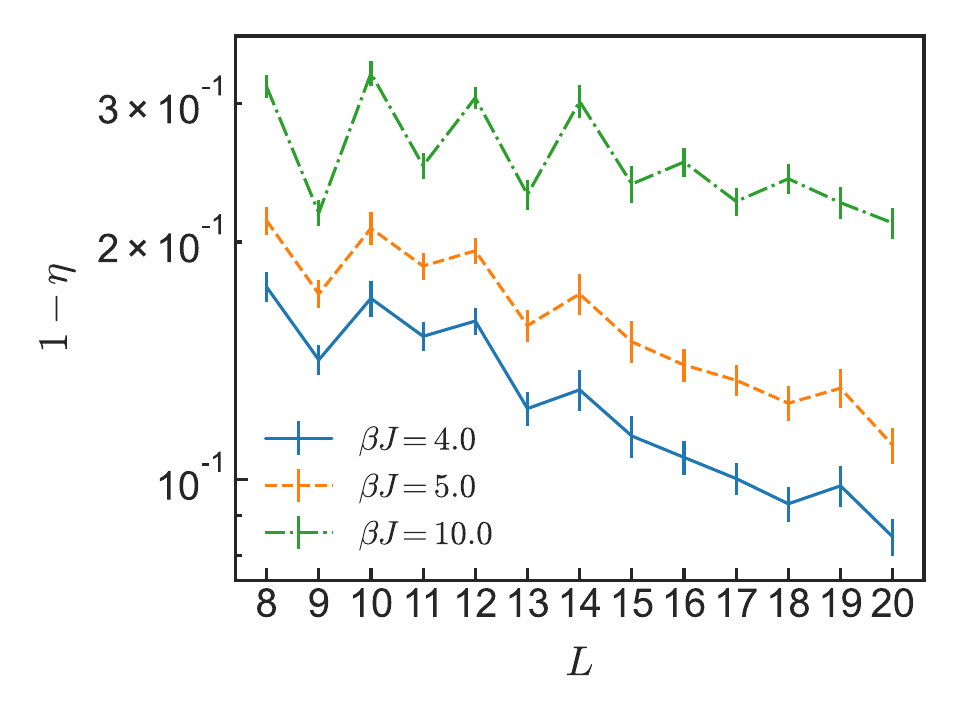}
    \caption{System-size dependence of the difference of the efficiency from unity for some temperatures. The efficiencies are obtained from 1024 samples for the RPPSs evolved by the Trotter gates obtained from the XXZ chain with the staggered magnetic field \(h=J\). The error bars indicate 1\(\sigma\) uncertainty estimated from the bootstrap analysis with 4000 resampled data.\label{fig:eff_beta}}
\end{figure}

Figure~\ref{fig:eff_log} shows the system-size dependence of the difference of the efficiency from unity.
The shown data suggest that the difference decreases exponentially with the system size for both efficiency-increasing cases.
Figure~\ref{fig:eff_beta} indicates that similar decrease can be observed at other temperatures.
According to the error propagation formula
\begin{align}
    \label{eq:error}
    \mathrm{Var} \left( \frac{\overline{x}}{\overline{y}} \right) \approx \left( \frac{\overline{x}}{\overline{y}} \right)^2 \left[ \frac{\mathrm{Var} \ x}{\overline{x}^2} + \frac{\mathrm{Var} \ y}{\overline{y}^2} - 2  \frac{\mathrm{Cov}(x, y)}{\overline{x} \ \overline{y}}\right]
\end{align}
and the discussion in \textcite{iwaki_purity_2022-1}, the variance of sampled thermal expectation values can be bounded with some constant  \(D\) as
\begin{align}
    \mathrm{Var} \braket{\hat{O}}_\beta \lesssim D \|\hat{O}\|^2 \ln \left(\frac{1}{\eta}\right)
\end{align}
when the efficiency is sufficiently close to unity.
With the efficiency represented by \(\eta = 1 - \delta(L)\), the upper bound is given as
\begin{align}
    D \|\hat{O}\|^2 \ln \left(\frac{1}{\eta}\right) &= -D \|\hat{O}\|^2 \ln \left(1 - \delta(L)\right) \nonumber \\
    &\simeq D\|\hat{O}\|^2 \delta(L)
\end{align}
when \(\delta(L) \ll 1\).
Therefore, the variances of the sampled expectation values decrease with the system size as rapidly as or more rapidly than \(1-\eta \).
\begin{figure}
    \includegraphics[width=0.95\linewidth]{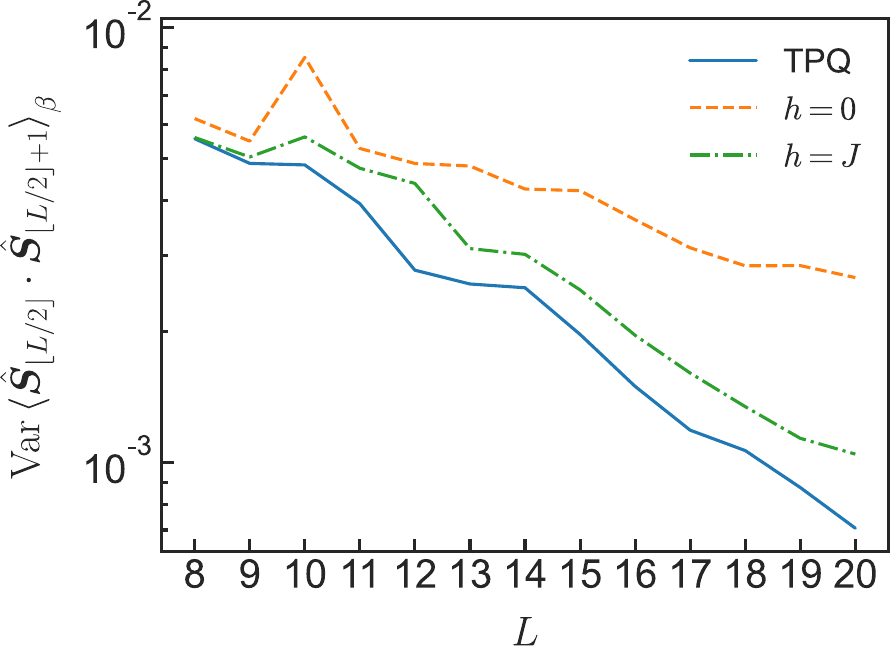}
    \caption{System-size dependence of the variance of neighboring spin correlation at the center of the system. The variance is estimated from the error propagation formula~\eqref{eq:error}. The settings of simulations are the same with those of Fig.~\ref{fig:HeisXXZ}.\label{fig:nn_center}}
\end{figure}
For instance, the system dependence of the variance of neighboring spin correlation at the center of the system is presented in Fig.~\ref{fig:nn_center}.
As expected from the exponential approach of the efficiency to unity, the variance decreases exponentially with the system size.
This fact indicates that the number of samples required for the statistical error of a sampled expectation value to be below a certain threshold exponentially decrease as well.

\begin{figure}
    \includegraphics[width=0.95\linewidth]{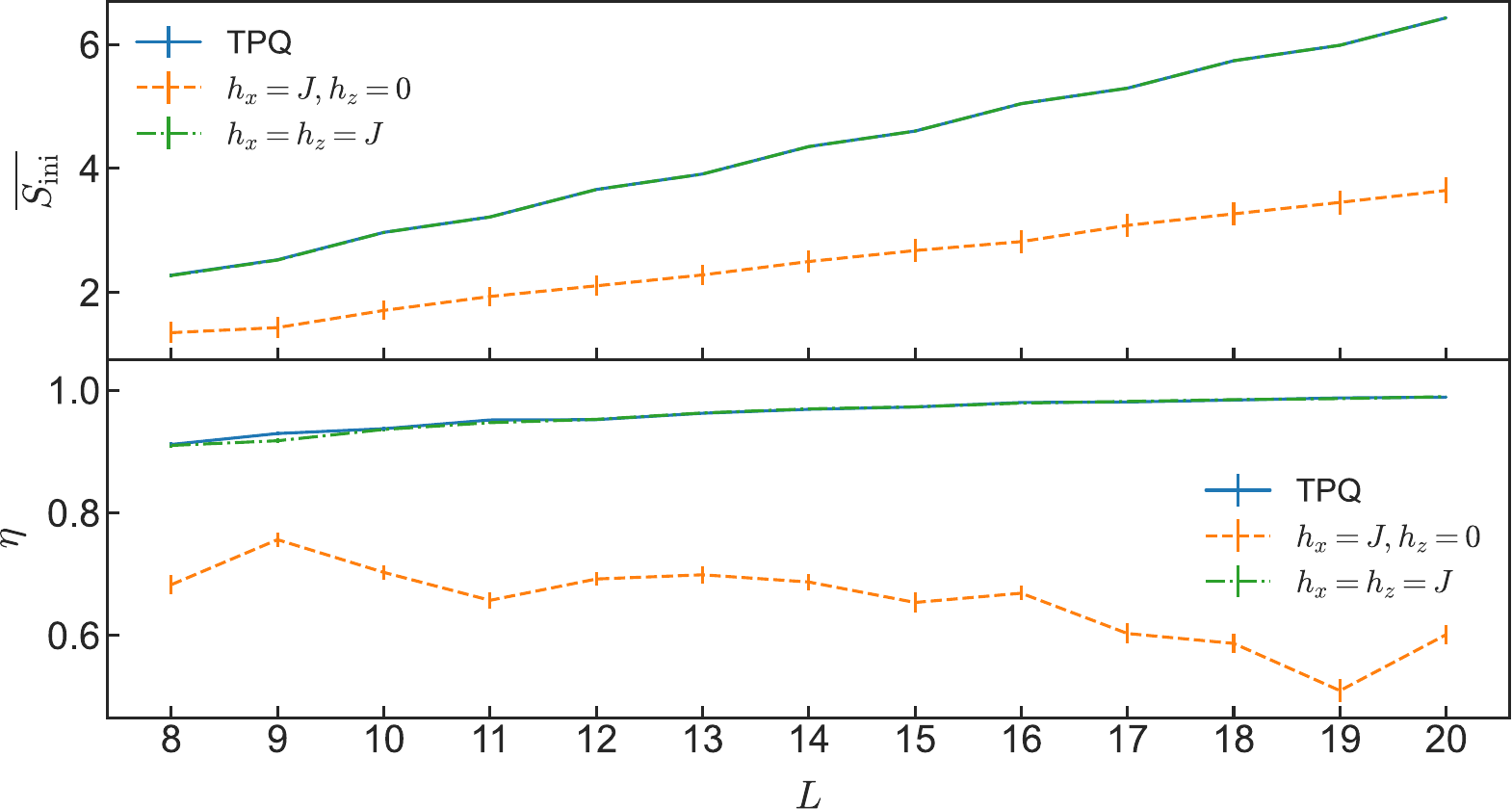}
    \caption{System-size dependence of the average entanglement entropy of sampled initial states (top panel) and the efficiency (bottom panel) obtained from 1024 samples for the TPQ states (blue solid line), RPPSs evolved by the Trotter gates obtained from the transverse-field Ising chain (orange dashed line), and those evolved by the Trotter gates obtained from the mixed-field Ising chain (green dashed-dotted line). The inverse temperature \(\beta \) is set to \(3.0 J^{-1}\). The error bars indicate 1\(\sigma \) uncertainty. The 1\(\sigma \) uncertainty of the efficiency is estimated from the bootstrap analysis with 4000 resampled data.\label{fig:HeisIsing}}
\end{figure}

In order to add another case that supports our conjecture, we analyze another Trotter Hamiltonian,
\begin{align}
    \hat{H}_\mathrm{T} = J \sum^{L-1}_{i=1} \hat{S}^z_i \hat{S}^z_{i+1} + \sum^L_{i=1}(h_z \hat{S}^z_i + h_x\hat{S}^x_i).
\end{align}
Here, \(h_x\) (\(h_z\)) is the magnetic field along the \(x\)- (\(z\)-) direction.
This Hamiltonian is often called the transverse-field Ising chain when only \(h_x\) is finite.
When both \(h_x\) and \(h_z\) are finite, the Hamiltonian is called the mixed-field Ising chain.
The transverse-field (mixed-field) Ising chain is integrable (nonintegrable)~\cite{banuls_strong_2011}.
Figure~\ref{fig:HeisIsing} represents the system-size dependencies of the averaged initial entanglement entropy and the efficiency that is estimated from 1024 samples obtained by the TPQ methods and the RPPSs that are evolved by the Trotter gates made from the transverse-field and the mixed-filed Ising chains.
Like the above XXZ chain case, the efficiency in the case of the nonintegrable mixed-field Ising chain increases with the system size and such an increase is absent in the case of the integrable transverse-field Ising chain.
Moreover, the efficiency of the mixed-field Ising chain is almost equal to that of the TPQ state method: Ideal sampling is achieved.

\begin{figure}
    \includegraphics[width=0.95\linewidth]{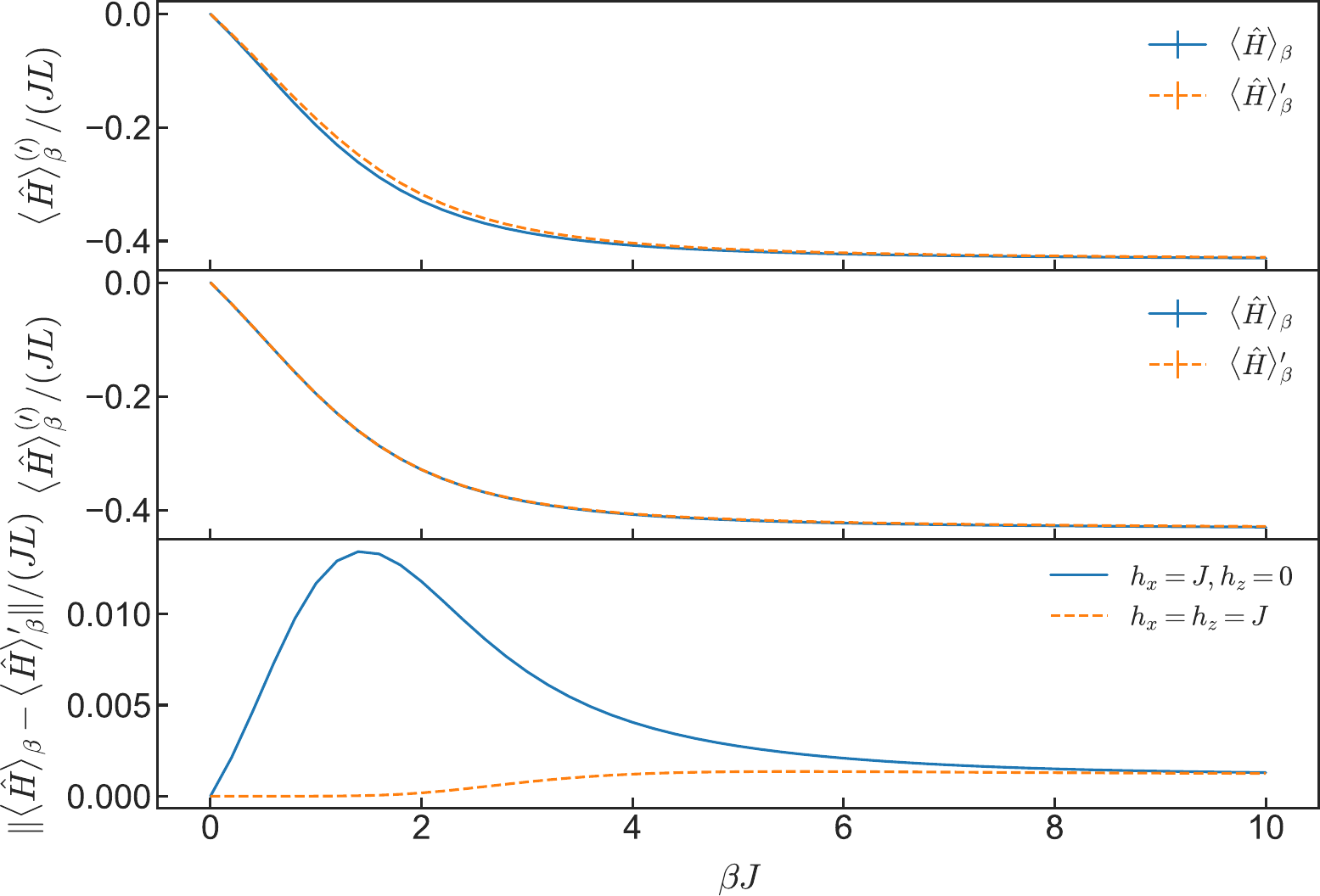}
    \caption{Inverse-temperature dependence of the internal energies of the Heisenberg chain estimated from Eqs.~\eqref{eq:th_exp_sample} and~\eqref{eq:th_exp_simple}. The top (middle) panel corresponds to the simulation with the Trotter gates obtained from the transverse-field (mixed-field) Ising chain with \(h_x = J\) (\(h_x = h_z = J\)). The system size \(L\) is set to 20. Error bars indicate 1\(\sigma \) uncertainty, and they are on the order of \(10^{-4}\) and thus smaller than the line width. The differences of the internal energies per site estimated from the two equations are given in the bottom panel.\label{fig:WoNorm}}
\end{figure}

The efficiency in the case of the mixed-field Ising chain is estimated to be 0.99 with \((L, \beta J) = (20, 3.0)\).
With such high efficiency, the replacement of \(\braket{\hat{O}}_\beta \)~\eqref{eq:th_exp_sample} by \(\braket{\hat{O}}^\prime_\beta \)~\eqref{eq:th_exp_simple} for the estimation of a thermal expectation value would be justified.
Figure~\ref{fig:WoNorm} shows the comparison between internal energies \(\braket{\hat{H}}_\beta \) estimated from Eq.~\eqref{eq:th_exp_sample} and \(\braket{\hat{H}}^\prime_\beta \) from Eq.~\eqref{eq:th_exp_simple}.
The system size \(L\) is set to 20.
Even in the case of the integrable transverse-field Ising chain, internal energies estimated from both equations show good agreement except visible difference around \(\beta J = 2.0\).
With the nonintegrable mixed-field Ising chain, the differences of internal energies are much smaller and on order of \(10^{-3}J\) per site.
This means that the evaluation of the norms of imaginary-time evolved states can be skipped even in classically simulatable scale.

\begin{figure}
    \includegraphics[width=0.95\linewidth]{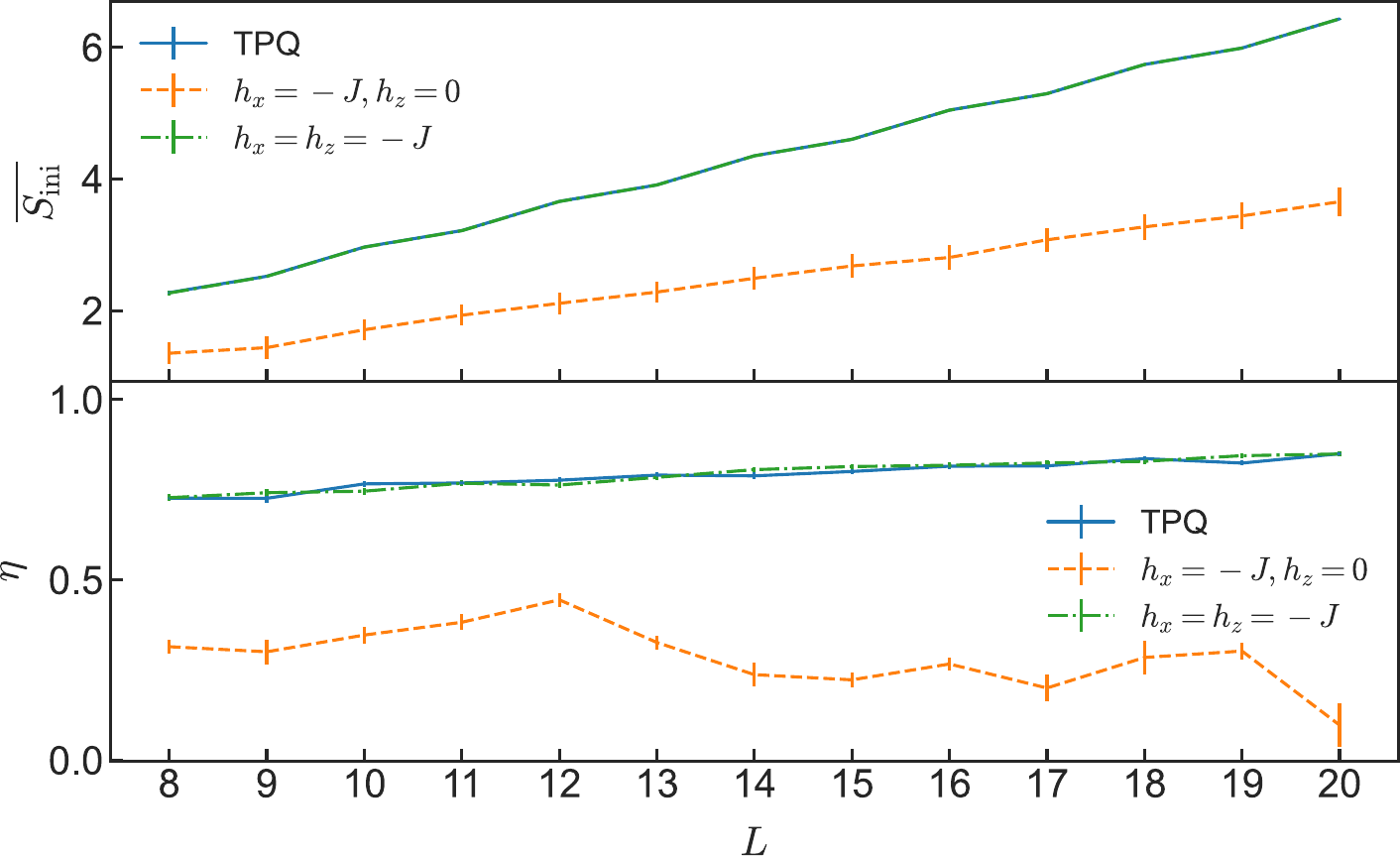}
    \caption{System-size dependence of the average entanglement entropy of sampled initial states (top panel) and the efficiency (bottom panel) obtained from 1024 samples for the TPQ states (blue solid line), RPPSs evolved by the Trotter gates obtained from the transverse-field Ising chain (orange dashed line), and those evolved by the Trotter gates obtained from the mixed-field Ising chain (green dashed-dotted line). The system Hamiltonian is the mixed-field Ising chain in this simulation. Inverse temperature \(\beta \) is set to \(3.0 J^{-1}\). The error bars indicate 1\(\sigma \) uncertainty. The 1\(\sigma \) uncertainty of the efficiency is estimated from the bootstrap analysis with 4000 resampled data.\label{fig:IsingIsing}}
\end{figure}

We also consider another system Hamiltonian
\begin{align}
    \hat{H} = J \sum^{L-1}_{i=1} \hat{S}^z_i \hat{S}^z_{i+1} + J\sum^L_{i=1}(\hat{S}^z_i + \hat{S}^x_i),
\end{align}
i.e., the mixed-field Ising chain.
The Trotter Hamiltonian is also set to the transverse-field or mixed-field Ising chain whose magnetic fields are chosen to be different from those of the system Hamiltonian.
The system-size dependencies of the efficiency are given in Fig.~\ref{fig:IsingIsing}.
Also in this case, the Trotter gates made from the mixed-field Ising chain makes the efficiency almost equal to that of the TPQ state method and the increase of the efficiency is absent with the Trotter gates made from the transverse-field Ising chain.

\section{Discussion and summary\label{sec:summaries}}

We have observed that the integrability of the Trotter Hamiltonian strongly affects how the efficiency of sampling depends on the system size. 
The integrability is closely related to chaoticity of the dynamics induced by the Trotter gates.
Under an integrable Hamiltonian, an extensive number of integrals of motion determined by initial states are conserved during real-time dynamics.
In other words, time-evolved states cannot forget initial memory and thus the dynamics is far from being chaotic~\cite{vidmar_generalized_2016}.

Comparing the two nonintegrable models that we have considered in this paper, namely the XXZ chain with the staggered magnetic field and the mixed-field Ising model, the latter shows better performance which is comparable to that of the TPQ state method.
In terms of conserved quantities, the XXZ chain with the staggered field conserves the magnetization in \(z\)-direction while the mixed-field Ising chain does not.
In other words, the Trotter gates made from the mixed-field Ising chain induce transitions between states with different magnetization while such transitions are absent when the Trotter gates are made from the XXZ chain with the staggered field.
This difference would result in the difference of the performance, and thus a Hamiltonian with less conserved quantities can be more suited for a Trotter Hamiltonian.

\begin{figure}
    \includegraphics[width=0.95\linewidth]{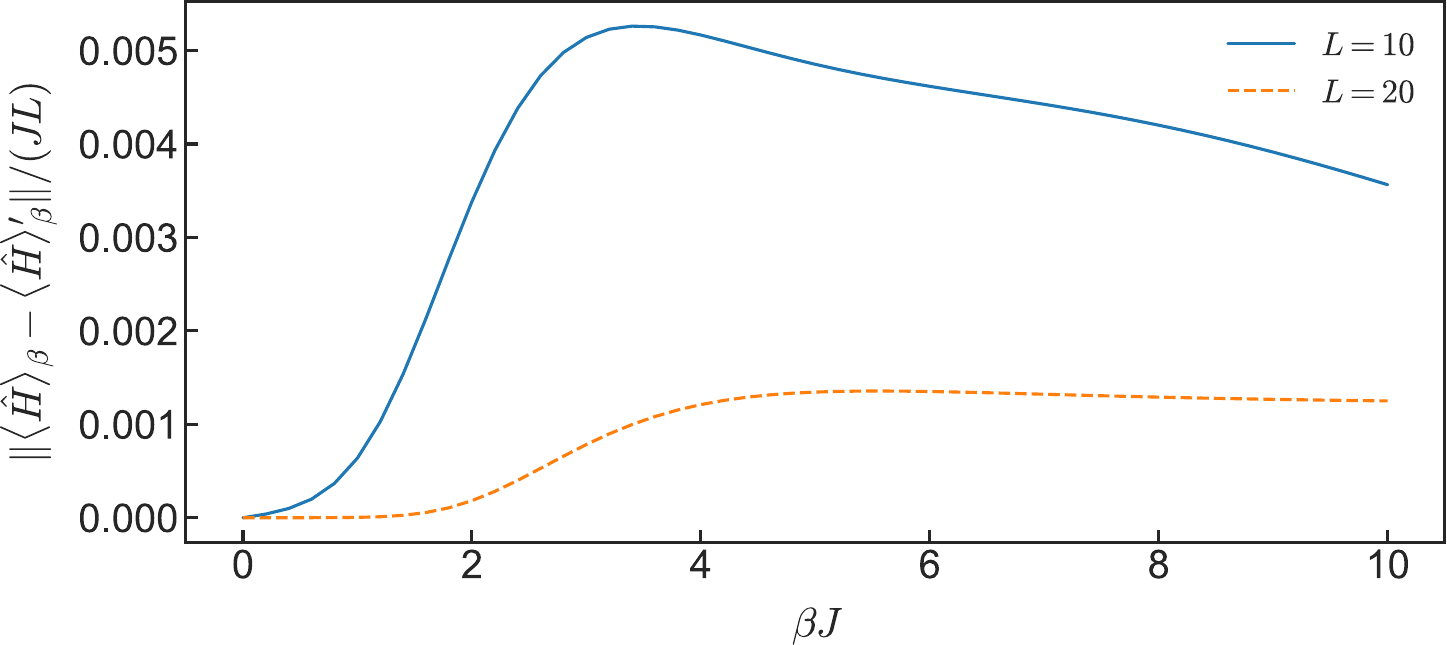}
    \caption{Differences of the internal energies per site estimated from Eqs.~\eqref{eq:th_exp_sample} and~\eqref{eq:th_exp_simple} with different system sizes \(L=10\) and \(20\). The system Hamiltonian and the Trotter Hamiltonian are set to the Heisenberg chain and the mixed-field Ising chain with \((h_x, h_z) = (J, J)\), respectively.\label{fig:compL}}
\end{figure}

Estimating the efficiency in quantum simulation would be troublesome because the norms of imaginary-time evolved states, which are required for the estimation, depend exponentially on the inverse temperature.
Instead of estimating the efficiency, it would be useful to evaluate the performance of nonintegrable Trotter gates implemented in quantum circuits by comparing the expectation values obtained by classical simulations and those estimated from Eq.~\eqref{eq:th_exp_simple} by quantum simulation in small systems.
If the expectation values obtained from classical and quantum simulations show good agreement, this agreement is expected to remain in larger systems because the efficiency increases with system size.
Figure~\ref{fig:compL}, which shows the differences of internal energies per site estimated from Eqs.~\eqref{eq:th_exp_sample} and~\eqref{eq:th_exp_simple} with different system sizes, supports this expectation.
Consequently, one can roughly verify the performance of Trotter gates without accessing the norms of imaginary-time evolved states.

Another candidate approach for evaluating thermal expectation values is the minimally entangled typical thermal states (METTS) approach which is designed for matrix-product-state simulations~\cite{white_minimally_2009,stoudenmire_minimally_2010}.
The METTS approach has the importance sampling scheme implemented via the Markov chain, and the autocorrelation of the Markov chain can be significantly relaxed~\cite{white_minimally_2009,stoudenmire_minimally_2010,goto_minimally_2020}.
Thus, the sampling problem discussed in this paper is absent.
The execution of the METTS approach on quantum devices has already been conducted~\cite{sun_quantum_2021}.

The advantage of our proposed method over the METTS approach appears when the almost ideal sampling is achieved with the proposed method.
In the almost ideal sampling, a single pure state possesses the information of finite-temperature quantities.
Therefore, one can evaluate the expectation values of observables from the single pure state. 
On the contrary, in the METTS approach, one has to evaluate the expectation values obtained from states appearing in the Markov chain.
In the proposed method, the number of pure states to be investigated can be much smaller than that in the METTS approach.

In summary, we proposed a method of initial-state sampling for the estimation of a thermal expectation value of quantum many-body systems and evaluated the system-size dependence of its efficiency.
In our proposed method, initial states can be generated from product states by applying Trotter gates \(2L\) times, where \(L\) is the system size.
We performed numerical simulations of several specific models on classical computers in order to show that the efficiency increases with system size when the Trotter gates are made from a nonintegrable Hamiltonian.
This increase is absent when the Trotter gates are made from an integrable Hamiltonian.
When the mixed-field Ising chain is adopted as the Trotter Hamiltonian, the efficiencies in studied models are almost equal to those of the thermal pure quantum states method, which can be regarded as an ideal sampling.
Our observations would contribute to construct an efficient quantum algorithm to estimate thermal expectation values of quantum many-body systems.

\begin{acknowledgments}
We thank A.\ Iwaki, C.\ Hotta, T.\ Okubo, M.\ Kunimi, and L.\ Coopmans for fruitful discussions and comments.
This work was financially supported by JSPS KAKENHI (Grant No.\ 18H05228, No.\ 20K14377, No.\ 21H01014, and No.\ 21K13855), by JST FOREST (Grant No.\ JPMJFR202T), and by MEXT Q-LEAP (Grant No.\ JPMXS0118069021).
\end{acknowledgments}
\input{main.bbl}
\end{document}

%% file: main.bbl
%